\documentstyle[titlepage,twoside,fleqn,12pt,epsf]{article}

\def\ref#1{\par\noindent \hangindent=0.4in \hangafter=1 #1 \par}
\def\eqalign#1{\null\,\vcenter{\openup\jot \m@th
  \ialign{\strut\hfill$\displaystyle{##}$&$
     \displaystyle{{}##}$\hfill \crcr#1\crcr}}\,}
\def\tempest%
{\begin{array}{ccc}
1 & 1 & 1 \\
1 & 1 & 1 \\
4 & 3 & 8
\end{array}}

\begin{document}
\bigskip
\centerline{\Large Luminosity Function of the Perigalactocentric Region}
\bigskip
\centerline{\bf Vijay K. Narayanan}
\centerline{\bf Andrew Gould \footnote{Alfred P.\ Sloan Foundation Fellow}}
\centerline{\bf D. L. DePoy}
\bigskip
\centerline{Dept.\ of Astronomy, The Ohio State University, Columbus, OH 43210}
\smallskip
\centerline{ vijay@payne.mps.ohio-state.edu}
\centerline{ gould@payne.mps.ohio-state.edu}
\centerline{ depoy@payne.mps.ohio-state.edu}
\bigskip
\bigskip

\centerline{\bf Abstract}
We present $H$ and $K$ photometry of $\sim 42,000$ stars
in an area of $\sim 250$ arcmin$^{2}$ centered on the Galactic center.
We use the photometry to construct a dereddened $K$ band
luminosity function (LF) for this region, excluding the excessively crowded
inner $2'$ of the Galaxy.
This LF is intermediate between the LF
of Baade's window and the LF of inner $2'$ of the Galactic center.
We speculate that the bright stars in this region have an age which is
intermediate between the starburst population in the Galactic center
and the old bulge population.
We present the coordinates and mags for 16 stars with $K_{0} \leq 5$ for
spectroscopic follow up.

\bigskip
\noindent
{\it Subject Headings}: Galaxy: center-infrared: stars-stars: luminosity
function - ISM: dust, extinction

\bigskip
\centerline{submitted to {\it The Astrophysical Journal}: March 12, 1996}
\centerline{Preprint: OSU-TA-4/96}

\newpage

\section{Introduction}
The $2.2 \mu $m luminosity functions (LFs) of Baade's window (BW) and the
inner $2'$ of the Galactic center (GC) are markedly different
at the bright end.
The GC LF has a substantially higher fraction of stars brighter
than $K_{0} = 5.5$, but it is nearly identical to the BW LF at the faint
end for  $K_{0} >7.0$.
This suggests that the GC contains a population of younger stars
in addition to an older population of red giants similar to
the one in BW (Haller 1992; Blum et al. 1996).
Here, we present $H$ and $K$ photometry of  $\sim 42,000$ stars in an
area $\sim 16' \times 16'$ centered on the GC.
In our analysis, we exclude the inner $2'$ and refer to the remaining
area as the Perigalactocentric region (PGC).
We construct the $K$ band LF of the PGC by individually
dereddening each star using its observed $(H-K)$ colors.
We find that the PGC does not contain any super bright stars
of the type found in the GC.
For $K_{0} < 6.0$, the GC LF has a significant excess over the PGC
LF.
In comparing the PGC with BW, we find that for  BW the
contamination by the foreground distribution of $M$ giants is significant
for $K_{0} < 7.0$, while for the PGC it is negligible.
Even if the contamination in BW is neglected, there is a still a clear excess
of stars in the PGC over BW in the range $4.0 < K_{0} < 7.0$.
We conclude that the luminosity function  of the PGC is intermediate
between that of the GC and Baade's window.

\section{Observations and Mosaic construction}

\par

We constructed mosaic images of the GC and PGC  in $H$ and
$K$ bands by stitching together $8 \times 19 = 152$ overlapping images.
The grid of images was taken on UT 1995  June $10$ and $11$
using the Ohio State Infrared Imaging Spectrograph (OSIRIS; DePoy et al. 1993)
at the $1.8$m Perkins Telescope, located on Anderson Mesa near
Flagstaff, Arizona.
OSIRIS uses a $256 \times 256$ NICMOS III array.
All the images were taken under photometric conditions.
The plate scale is $\sim 0.\hskip-2pt''63$ per pixel $(\sim 155''$ field
of view).
Images in the same row are offset by $\sim 50''$ while those in the
same column are offset by  $\sim 120''$.
The images were exposed successively without any
intervening sky exposures.
The seeing was between $1.\hskip-2pt''4$ and $2.\hskip-2pt''4$
in $H$ and between $2.\hskip-2pt''1$ and $2.\hskip-2pt''8$ in $K$.
We constructed the $H$ and $K$ mosaics independently.
A standard star P565-C  was observed  both before and after
the observations in the $H$ band and after
the observations in the $K$ band.
Each exposure was $\sim 1.1 $ seconds long.
We coadded $10$ individual exposures to form each individual image.
The sky image formed from the median of the standard star images
was subtracted from each of the raw images.
Finally, we divided the images by a flatfield formed from the
median of the images of a white screen mounted to the telescope.

\par

We aligned the images geometrically and photometrically,
before making the final mosaic.
We cross-correlated every pair of overlapping images
in order to determine $4$ parameters that relate the images$\colon$
a vertical offset, a horizontal offset, a sky offset,
and the relative intensity normalization (due to differences in
atmospheric transparency and exposure times).
We obtained a global solution for these parameters by
weighting each pair according to the number of overlapping
pixels.
Bad pixels and saturated stars were rejected before the cross
correlation.
The overlapping sections of the images were shifted by fractions of
a pixel using linear interpolation.

\par

The global solution for the geometric and photometric offsets
between the images was used to construct the mosaic
in the $H$ and $K$ bands.
In combining the pixels, we averaged all the contributing
pixel values, after rejecting the  pixel values that were
more than $5 \rm{\sigma}$ away from the mean of the
remaining pixel values, where $\rm {\sigma}$ is computed
assuming photon counting statistics.
In the case of bright stars, we accepted all the contributing
pixel values to ensure that the flux of the star is conserved.
This procedure does not eliminate cosmic ray events
when they lie on top of bright stars.
However the short exposure times of the images should ensure that
the cosmic ray events do not significantly contaminate the image.
We verified this by visually scanning the image.
Some of the individual $H$ band images contain interference
fringes from variable OH emission with typical
amplitude of $\sim 6$ ADU and width of $\sim 15 $ pixels.
Fringes of this type are often removed from $H$ band images
by chopping to the sky.
As noted above, we did not chop.
However, the typical amplitude of the fringes in the
mosaic image is only $\sim 2$ ADU (since typically,
each pixel in the final mosaic is an average of pixels
in 3 individual images), corresponding to
H $\sim 19$ mag arcsec$^{-2}$ and hence too small
to affect the results.

\par

The $H$ and $K$ mosaics are shown in Figures 1 and 2 respectively.
The $K$ mosaic is offset by $\sim 40''$ to the west relative
to the $H$ mosaic.
Excluding the rim of the mosaic where there is information
from only one image, about $60\% $ area of each
mosaic is constructed from $3$ overlapping images and
the other $40\% $ from $6$ images.
The typical sky level is $\sim 14$ mag arcsec$^{-2}$ in the $H$ mosaic
and $\sim 12$ mag arcsec$^{2}$ in the $K$ mosaic.
One ADU corresponds to $\sim 21$ mag arcsec$^{-2}$ in $H$ and
$\sim 20.4$ mag arcsec$^{-2}$ in $K$.
There is a difference of $\sim 2$ to $3$ ADU across the
seams in the final mosaic, of order $1\%$ of the sky.

\section{Photometry}

\par

We initially identified $\sim 28,000$ candidate stars in the $H$ mosaic
using the DAOFIND routine of DAOPHOT (Stetson 1987).
The extinction pattern in the images has many patchy and filamentary
structures on small scales.
For stars in these highly obscured patches, the local sky determined
from an annular region  around the star by DAOPHOT was often greater than
the peak flux of the star itself.
These stars were either not identified or photometered incorrectly
by DAOPHOT.
We therefore  manually identified the stars not found by DAOFIND
by displaying the $H$ mosaic in various stretches of brightness levels.
We used the $K$ mosaic  to help identify the
faint stars in some heavily obscured regions.
However, the sample of stars is essentially $H$-band selected,
because of better seeing in $H$.
Finally, we identified a total of $\sim 65,000$ candidate stars.

\par

The photometry was carried out by extracting small,
overlapping sections of $100 \times 100$ pixels from the $H$ and $K$
band mosaics.
The overlap between the neighboring sections is 5 pixels
in both the horizontal and vertical directions.
There are 342 such sections in each band.
Each of these sections was photometered independently.
In each section, only the stars that lie within the inner $95 \times 95$
pixels were accepted in the final photometry list.
We measured the $H$ and $K$ magnitudes of the stars by fitting
the stellar profiles to an appropriate point spread function (PSF) and the
local sky.
The instrumental magnitudes determined from this PSF fitting
were converted to apparent magnitudes using aperture photometry
of PSF stars compared to the standard stars.

\par

The observations were carried out at $\sim 2$ air masses, while
the standard star was observed at $\sim 1$ air mass.
We therefore applied an air mass correction of 0.1 mag in both
$H$ and $K$ bands.
There is a residual uncertainty of $\sim 0.03 $ mag in
the photometric zero points because of the uncertainty in the
appropriate value of this correction.

\subsection{$H$ band}
\par

The seeing in $H$ band changed significantly during the approximately 2 hour
period of observations.
Therefore, we identified 98 stars that were well isolated and
bright enough that they were not contaminated by the light from
neighboring stars.
These PSF stars are spread approximately  uniformly throughout the $H$ image.
The stars in each of the 342 small sections were photometered using
a PSF that was most similar to the PSF  for that section.
Except in a few cases where the seeing changed abruptly,
the PSF nearest to a section was adequate for this purpose.

\par

We fit all the stars in a section simultaneously.
Starting from the approximate coordinate locations, we first fit
each star to $4$ parameters - $3$ parameters that model the sky (including
a constant offset  and linear terms in the horizontal and vertical directions)
and $1$ parameter for the PSF normalization.
With these parameters, we then find the best values for the coordinate
locations of the star by shifting the stellar profile by fractional
pixel amounts and minimizing the $\chi^{2}$ difference between
the fit and the image.
This best fit for the star is then subtracted from the image.
All the stars are fitted and subtracted from the original image
in this manner to create a residual image.

\par

The brightest star was then added to this residual image.
It was again fit for all the $6$ parameters that define the
PSF normalization, the local sky, and the best coordinate locations
in the same manner as described above, and this best fit was again
subtracted from the residual image.
This procedure was repeated for all the other stars in order of
their decreasing brightness as determined from the first pass.
Therefore, before we fit for any star, all its brighter neighbors would have
already been measured and their fluxes subtracted, thereby reducing
the contamination from  neighbors.

\subsection{$K$ band}
\par

The $K$ mosaic is distorted by $\sim 0.1\%$ with respect to the $H$ mosaic.
We determined a linear transformation between the $H$ and $K$ mosaics,
using the locations of $142$ bright stars.
The distortion was largely removed by this transformation, as
was verified by the residuals of the fit.
The seeing in $K$ band is worse than the $H$ band.
There are only a few isolated stars that are not contaminated by
neighboring stars.
However, the seeing remained approximately constant throughout the
$K$ band observations.
Therefore, all the stars in the $K$ band were fit to $5$ different
PSFs in a manner similar to the $H$ photometry.
In this fitting procedure however, we assumed that the transformed
locations of the stars are accurate enough.
For each star photometered in $H$, we assumed that the $K$ mosaic also
contained the star at the corresponding transformed location.
Therefore, each star was fitted for $4$ parameters only.
We take the average of these $5$ measurements (after rejecting $3 \sigma$
outliers) to be the $K$ mag.
We estimate the error in the $K$ photometry as the standard deviation
of the $5$ measurements averaged over the stars in a given mag bin
(Fig. 3).

\subsection{Errors in the Photometry}
\par
$H$ and $K$ mag are measured for $\sim 42,300$
stars in the entire field.
The other candidates are either misidentifications or
are too faint in one of the bands to be measured by this method.
The errors in $H$ band are assumed to be the same as the errors
in $K$ band.

\par
The photometric errors in both bands depend on the crowding of stars.
Very close to the GC, the photometry is
limited by our ability to identify separate stars.
Excessive crowding of stars, together with the relatively poor seeing
made it difficult to separate the objects into
individual stars.
The  results reported here  are based on
photometry of regions more than $2'$ away from the GC.

\section{Dereddened magnitudes}

\par

The observed $(H-K)$ color of a PGC star is the sum
of the intrinsic color $(H-K)_{0}$ and the
selective extinction coefficient, $E(H-K)$.
Therefore, we deredden the PGC stars individually using the
observed $(H-K)$ colors and an appropriate color-mag relation.
We use the near infrared extinction law determined by Mathis (1990)
to compute the total extinction coefficients $A_{K}$ and $A_{H}$.
Specifically, we assume $A_{H}/A_{K}= 1.66$ and
$A_{K}=1.5E(H-K)$.

\par

We assume that the intrinsic color-mag relation for
 stars in PGC (corrected for extinction) is the same
as that of stars in  BW.
We fit a color-mag relation for stars in
BW using the photometry provided to us by
J. Frogel (1995, private communication) to obtain the
following analytic relation,

$$K_{0} = -9(H-K)_{0}+10.7 \quad (\hbox{$K_{0} < 8.5$}),$$
$$K_{0} = -50(H-K)_{0}+18.0 \quad (\hbox{$K_{0} > 8.5$}),$$
$$\eqno{(4.1)}$$
All stars with $E(H-K) < 0.6$ are assumed to be foreground stars.
There are $10$ stars that are saturated either in the $H$ or $K$ mosaic,
or both.
Therefore, we divided the $H$ mosaic by the $K$ mosaic and determined the
colors of these stars by estimating the color at the wings,
where the pixel counts are not saturated.
We estimated the colors at the wings by comparing them with
nearby stars that were not saturated.
Only one of these 10 stars is sufficiently reddened to be a
PGC star.

\section{Luminosity function of the PGC}

\par

The PGC  suffers from very patchy extinction.
As a result,  the completeness level of our photometry can
change drastically throughout the imaged region.
Therefore, we identified $6$ regions around the GC
that have moderate ($A_{K} \sim 2.7$) and
roughly homogeneous extinction (collectively called Reg1).
While Reg1 does contain a few patches of high extinction,
these patches are very small and do not affect our results.
Reg1 encloses about $5\%$ of the total area of the PGC.
The $6$ regions comprising Reg1 are chosen at more than $2'$ away from the GC,
 so the photometry does not suffer from the severe crowding of the GC.
All the Reg1 stars  with $K_{0} \leq 8.0$
were examined individually and the bad photometry cases were rejected.
The color-mag diagram of patch 1 of Reg1 is shown in Figure 4.
In order to be able to compare the Reg1 LF  with
that of BW and GC, we normalize the star counts in a given region by the
integrated $K$ band flux (corrected for extinction) falling in that region.
We determine the dereddened flux as follows.
In each of the $6$ regions comprising Reg1,  we compute
 the mean $\bar E(H-K)$ of stars from the mean
offset in $(H-K)$ relative to the giants in BW.
We then find the total flux within the region and multiply this by
$10^{0.4\bar A_{K}}$, where $\bar A_{K}=1.5\bar E(H-K)$.
We estimate the true sky by measuring the counts in heavily extincted
regions.
This gives an upper limit to the true sky as there is some $K$ light
from the bulge even in heavily extincted regions.
This sky is then used to estimate the counts due to bulge light in
lightly extincted regions with a small fractional error due to the
uncertain sky.
By comparing the extinctions in these $2$ regions (measured from the reddening
of bright stars), we can accurately estimate the small amount of
$K$ light from the bulge in the heavily extincted region.
We estimate the errors in this procedure to be $\sim 3$ ADU per pixel.
This leads to an  uncertainty  of $\sim 2\%$ in the total flux, negligible
compared to other errors.

\par

To verify the accuracy of this LF of the PGC,
we also identified another $22$ regions of moderate reddening (Reg2),
covering another $20\%$ of the total area of the PGC.
All stars with $K_{0} \leq 6.5$ in these regions were
examined individually and the LF was constructed in a similar manner.
The LFs determined  for Reg1 and Reg2  are shown in Table 1
and are plotted in Figure 5.
The two LFs are identical within  Poisson errors.

\par

The $H$ and $K$ mosaics give the visual impression of a disk-like
feature approximately aligned with the Galactic plane (Figs. 1 and 2).
To test for the reality of this feature, we selected all the
Reg1 regions from within the apparent disk and all the Reg2 regions
from outside it.
The centroids and areas of the $6$ patches comprising Reg1 and the $22$
patches comprising Reg2 are in Table 2 and Table 3 respectively.
The locations and geometries of these patches are shown in Figure 6.
Although a few patches lie close to each other, they are reddened
by different amounts.
The LFs of these two regions are the same, indicating that the stellar
populations at least of these two regions are indistinguishable.

\section{Completeness limit of the LF}

\par

The LF is based on stars selected in the $H$ mosaic.
Stars were identified first by using the DAOPHOT star finding routine
DAOFIND and then by scanning the images manually
to find the stars missed by DAOFIND.
Based on the tests described below, we conclude that the
LF is complete up to $K_{0} = 8$.

\par

We added $10$ artificial stars per half mag bin in $H$ at random locations
in each patch.
We ran DAOFIND on the resulting images to determine the fraction
of stars detected.
For each undetected star, we judged whether we would have found the star
in the manual search.
Figure 7 shows the completeness of the DAOFIND and combined
(DAOFIND + manual) searches.
The combined search is complete for $H < 12.5$ and $\sim 80 \%$ complete
at $H = 14$.

\par

The DAOFIND completeness limit is objective, but the
DAOFIND $+$ manual search limit depends in part on subjective judgement.
To test whether this procedure reproduces the actual manual detection rate,
we compared the
fraction of stars that were undetected by DAOFIND but were found manually in
the real data, with the same fraction in the artificial-star sample.
The results of this test are shown in Figure 8.
The sample star list fractions lie within $1 \sigma$ of the artificial
star list fractions for all $H$.
This consistency check confirms the accuracy of the combined
(DAOFIND + manual) detection efficiency.

\par

To construct the LF, we weighted each detected star by the inverse
of the completeness fraction corresponding to its $H$ mag.
The patches typically have $A_{H} \sim 4.5$.
Therefore, the LF is fully complete for $K_{0} < 8$.
Due to the weighting procedure, we also expect the LF to be
statistically complete for $8 < K_{0} < 9$.

\par

Some of the stars in the  sample star list are bad identifications.
For $K_{0} < 8$, we found these by checking each star individually.
For $K_{0} > 8$, we applied the following statistical corrections.
We selected $100$ stars randomly in each of the mag ranges
$8 < K_{0} < 9$, $9 < K_{0} < 10$ and $10 < K_{0} <11$.
These stars were examined individually and bad identifications were rejected.
In these $3$ mag ranges, we found  $15 \%$, $22 \%$ and $22 \%$ of the total
sample of stars to be bad identifications.
For $K_{0} > 11$, we assumed the bad identifications to be $22 \%$.
We therefore multiplied the LF in these mag ranges by the fraction of good
identifications.

\section{Comparison to BW and GC LFs}

\par

We first compare the normalized $K$ band LF of the PGC
with that of BW (Fig. 9).
For $K_{0} < 6.5$, we use the information from both Reg1 and Reg2
to improve the statistics.
The LF for $K_{0} >6.5$ is constructed from Reg1 only.
We use the  BW LF from Tiede et al. (1995), which is
complete in the range  $5.5 \leq K_{0} \leq 16$.
The bright end  ($5.5 \leq K_{0} \leq 9.0$) is derived by
Frogel \& Whitford (1987) (hereafter referred to as FW87), from
photometry of an unbiased subsample of the complete $M$ giant surveys
by Blanco et al. (1984) and Blanco (1986).
These surveys also included $10$ Long period variables (LPVs).
For $9.0 \leq K_{0} \leq 12.5$, the BW LF is based
on the data of DePoy et al. (1993), who
conclude that the $M$ giants selected from  the
grism surveys by FW87 include essentially all stars with $K_{0} \leq 10.0$,
and that the FW87 field toward BW does not contain any other luminous stars.
To compare the PGC and BW LFs, we normalize the latter to the total flux
as follows.
First, we compute the total flux of the stars in the LF which have
$M_{bol} < -1.2 $ (corresponding to $K_{0} < 11.0$ at a distance to the
GC of $R_{0} = 8.0$ kpc).
FW87 and Frogel (1988) conclude that this flux accounts
for $60\%$ of the total $K$ band light.
We therefore divide this flux by $0.6$ to obtain the total flux,
and use it to normalize the BW LF.
Figure 9 also shows the GC LF obtained by Blum et al. (1996), who
determined the relative normalization of the BW LF and GC LF.

\par
Table 1 gives the 3 LFs together with  the actual number of
stars in  each half mag bin on which the LFs are based.
Unlike the GC, the PGC LF does not extend  brighter than $K_{0} = 4.0$.
In fact, we did not find a single star with a dereddened mag
$K_{0} < 4.0$ in either Reg1 or Reg2.
In the entire field of the PGC,  there was
only one star in the range $3.5 < K_{0} < 4.0$, and no stars
brighter than $K_{0}=3.5$.
In the range $4.0 \leq K_{0} \leq 6.0$, the GC and PGC
LFs are based on 43 and 33 stars respectively.
If the PGC LF were the same as the GC LF, we would expect
$\sim 162 \pm 13$ stars in the range $4.0 \leq K_{0} \leq 6.0$ inside
Reg1 and Reg2, which is inconsistent with the actual number observed at the
$9 \rm{\sigma}$ level.
The individual half mag bins of the PGC LF are also deficient
relative to the GC LF for all $K_{0} < 6.0$.

\par

The BW LF brighter than $K_{0} = 7.0$ from
Tiede et al. (1995) is based on a very small number of stars.
Further, even these few bright stars with $K_{0} < 7.0$
could be foreground stars (Tiede et al. 1995).
We estimate the number of foreground stars expected in each half mag
bin for the grism survey of $M$ giants used by FW87
for two different models of the disk LF.
The FW87 field covers an area of $468\ \rm{arcmin}^{2}$.
The disk is assumed to have a radial scale length of $3$ kpc in
both the models (Kent et al. 1991).
The local density of different $M$ giants given by
Garwood \& Jones (1987) is scaled appropriately to derive the
density of these stars in any volume element along the line of sight
to BW.
In the first model, (GJ model) the disk scale height of all the $9$
different spectral classes of $M$ giants is taken to be $300$ pc
(Garwood \& Jones 1987).
In the second model, we assume a disk scale height
of $165$ pc for all the spectral classes (Kent et al. 1991).
The expected foreground contamination in the FW87 field for
observation cones extending to distances of $4$ kpc and $6$ kpc are
given in Table 4.
This table also contains the number of $M$ giants that are
actually observed by FW87 in each half mag interval.
For $K_{0} < 7.0$, the expected number of foreground $M$ giants
toward the BW in both models  is comparable to the
actual number of stars that are  observed in the FW87 field.
On the other hand, we find that the the contamination of the PGC
stars by foreground stars is negligible ($ < 0.15$ per half mag bin).
To understand this result, note that the total observed flux in the PGC
regions corresponds to $K \sim 13.4$
mag arcsec$^{-2}$, corresponding to a dereddened surface brightness of
$K \sim 10.7$ mag arcsec$^{-2}$.
This is about $70$ times brighter than BW with $K \sim 15.3$
mag arcsec$^{-2}$.
Thus, although the absolute surface density of foreground disk stars
is higher in the PGC than BW, the ratio of disk to bulge stars is much
smaller.
Even if all the $6$ stars brighter than $K_{0} = 7.0$ detected in the
BW grism surveys are taken to be bulge  stars, the PGC LF still contains
a significant excess over the BW LF when all the bins
in the range $5.5 < K < 7.0$ are combined.
The PGC LF is therefore different from both the LF of the inner $2'$ of the GC
and the BW LF.
It lies intermediate between the LFs of the GC and the BW for all
$K_{0} < 7.0$.

\section{Discussion}

\par

The stellar population of the GC has an excess of bright stars
compared to the  the older population of BW (Haller 1992).
The central pc of the Galaxy contains helium rich,
luminous, blue, emission-line stars and Wolf-Rayet stars
with estimated zero-age main sequence masses of up to
$\sim 100M_{\odot}$ (Allen, Hyland \& Hillier 1990; Krabbe et al. 1991;
Blum, Sellgren, \& DePoy 1995a; Blum, DePoy, \& Sellgren 1995b;
Libonate et al. 1995; Krabbe et al. 1995).
A plausible scenario is a burst of star formation in the GC
$\sim 10$Myr ago (Krabbe et al. 1995).
Krabbe et al. (1995) also conclude that the intermediate mass asymptotic
giant branch stars
were formed in another burst of star formation $\sim 100$Myr ago.

\par
In the PGC, we do not find stars that are as luminous as
the brightest stars in the GC.
Nevertheless, there is a significant excess of stars with
$K_{0} \leq 7$ over the older population of the BW.
This could imply the existence of a population of
stars that is significantly younger than the old bulge population.
The best way to confirm this hypothesis would be to take
spectra of the bright stars in the PGC.
To this end, we present in Table 5 a list of all stars
with $K_{0} < 5$ in the PGC region.

\section{Conclusion}

\par

We have constructed the K band LF of the perigalactocentric
region outside the inner $2'$ of the GC, by individually
dereddening every star using its observed $(H-K)$ colors.
This LF is complete for $K_{0} <8$.
Unlike the GC LF, this LF does not have any bright end ($K_{0} < 4.0$)
component.
There is also a deficiency in the PGC LF compared to the GC LF
for all $K_{0} \leq 6$.
However, the PGC LF has a significant excess over the BW LF
for $K_{0} \leq 7$.
We conclude that the PGC LF is intermediate between  the
GC and the BW LFs.

\par

We thank Bob Blum, Jay Frogel, Kris Sellgren and David Weinberg for
helpful comments and suggestions.
We also thank R. Bertram for his help with the observations.
Work by A.G was supported in part by NSF grant AST-94-20746.
OSIRIS was constructed with support from the NSF grants AST90-16112
and AST92-18449.

\newpage
\vskip50mm
\centerline{\bf REFERENCES}
\bigskip

\ref{Allen, D.\ A., Hyland, A.\ R., \& Hillier, D.\ J., 1990, MNRAS, 244, 706}
\ref{Blanco, V.\ M.\ 1986, AJ, 91, 290}
\ref{Blanco, V.\ M.,\ McCarthy, M.\ F., \& Blanco, B.\ M. 1984, AJ, 89, 636}
\ref{Blum, R.\ D., Sellgren, K., \& DePoy, D.\L.\ 1996, ApJ, submitted}
\ref{Blum, R.\ D., Sellgren, K., \& DePoy, D.\L.\ 1995a, ApJL, 440, L17}
\ref{Blum, R.\ D., DePoy, D.\L., \& Sellgren, K. 1995b, ApJ, 441, 603}
\ref{DePoy, D.\ L., Atwood, B., Byard, P., Frogel, J.\ A., \& O'Brien, T.,\
1993, in SPIE 1946, ``Infrared Detectors and Instrumentation,'' p.667}
\ref{DePoy, D.\ L., Terndrup, D.\ M., Frogel, J.\ A., Atwood, B., \& Blum, R.,
1993, AJ, 105, 2121}
\ref{Frogel, J.\ A.,\ 1988, ARA\&A, 26, 51}
\ref{Frogel, J.\ A., \& Whitford, A.\ E.\ 1987, ApJ, 320, 199}
\ref{Garwood, R., \& Jones, T.\ J.\ 1987, PASP, 99, 453}
\ref{Haller, J. 1992, Ph.D Thesis, University of Arizona,Tucson}
\ref{Kent, S.\ M., Dame, T.\ M., \& Fazio, G.\ 1991, ApJ, 378, 131}
\ref{Krabbe, A.,\ Genzel, R.,\ Drapatz, S.,\ \& Rotaciuc, V., 1991, ApJL, 382,
L19}
\ref{Krabbe, A., et al. 1995, ApJ, 447, L95}
\ref{Libonate, S.,\ Pipher, J.\ L., Forrest, W.\ J., \& Ashby, M.\ L.\ N. 1995,
ApJ, 439, 202}
\ref{Mathis, J.\ S.\ 1990, ARA\&A, 28, 37}
\ref{Stetson, P.\ B.,\ 1987, PASP, 99, 191}
\ref{Tiede, G., Frogel, J.\ A., \& Terndrup, D.\ M.\ 1995, AJ, 110, 2788}

\newpage
\setlength{\topmargin}{-0.7in}
\textwidth=6.5in
\oddsidemargin = 0.05in
\evensidemargin = 0.05in
\begin{table}
\begin{center}
\caption
{LFs and the number of stars actually observed in each half
magnitude bin for the $2$ sets of regions of PGC, the inner $2'$ of GC
and BW.
}

\bigskip
\begin{tabular}{*{9}{c}}
\hline\hline
$K_{0}$ & \multicolumn{4}{c}{PGC} & \multicolumn{2}{c}{GC} &
\multicolumn{2}{c}{BW$^{\dagger}$ }\\ \cline{2-5}
& \multicolumn{2}{c}{REG1} & \multicolumn{2}{c}{REG2} \\ \cline{2-3}\cline{4-5}
& LF & N & LF & N & LF & N & LF & N \\ \hline
1.75 & 0 & 0 & 0 & 0 & 0.46 & 1 & 0 & 0 \\
2.25 & 0 & 0 & 0 & 0 & 0.00 & 0 & 0 & 0 \\
2.75 & 0 & 0 & 0 & 0 & 0.91 & 2 & 0 & 0 \\
3.25 & 0 & 0 & 0 & 0 & 0.46 & 1 & 0 & 0 \\
3.75 & 0 & 0 & 0 & 0 &  1.83 & 4 & 0 & 0 \\
4.25 & 0.43 & 1 & 0 & 0 & 0.92 & 2 & 0 & 0 \\
4.75 & 0.43 & 1 & 0.54 & 4 & 3.66 & 8 & 0 & 0 \\
5.25 & 0.86 & 2 & 0.41 & 3 & 5.03 & 11 & 0 & 0  \\
5.75 & 3.47 & 8 & 1.90 & 14 & 10.05 & 22 & 2.56 & 1 \\
6.25 & 11.60 & 26 & 7.24$^{\ast}$ & 53 & 10.51 & 23 & 0.92 & 1  \\
6.75 & 20.25 & 46 & 20.96 & 179 & 34.29 & 75 & 7.01 & 4 \\
7.25 & 31.80 & 71 & 38.63 & 328 & 53.03 & 116 & 51.22 & 14 \\
7.75 & 42.36$^{\ast}$ & 94 & 58.50 & 495 & 64.10 & 139 & 52.12 & 15 \\
8.25 & 107.83 & 210 & 76.48 & 645 & 84.74  & 187 & 49.92 & 10 \\
8.75 & 161.09 & 274 & 91.67 & 772 & 97.83 & 214 & 104.77 & 14 \\
9.25 & 124.74 & 209 & 87.21 & 800 & 128.92 & 282 & 198.20 & 340 \\
9.75 & 133.63 & 215 & 99.84 & 916 & 146.29 & 320 & 287.50 & 505 \\
10.25 & 127.93 & 212 & 84.58 & 776 & 122.06 & 267 & 431.24 & 663 \\
10.75 & 156.64 & 196 & 80.88 & 742 & 65.37 & 143 & 574.98 & 925 \\
11.25 & 125.91 & 152 & 72.48 & 665 & 30.63 & 67 & 718.74 & 1309\\
11.75 & 90.31 & 112 & 59.85 & 549 & 13.25 & 29 & 1006.23 & 1786 \\ \hline
\end{tabular}
\vskip5mm
$\dagger$ For $5 \leq K_{0} \leq 9$, the photometry is from FW87.
For $9 \leq K_{0} \leq 12,$ the photometry is from DePoy et al. (1993).
\ \ \ \ \ \ \ \ \ \ \ \ \ \ \ \ \ \ \ \ \ \ \ \ \ \ \ \ \ \ \ \ \ \ \ \ \ \ \ \
\ \ \ \ \ \ \ \ \ \ \ \ \ \ \ \ \ \ \ \ \ \ \ \ \
\newline
$\ast$ Stars in this bin and brighter were examined individually. For fainter
mag, we applied statistical corrections.
For $9 \leq K_{0} \leq 12,$ the photometry is from DePoy et al. (1993).
\ \ \ \ \ \ \ \ \ \ \ \ \ \ \ \ \ \ \ \ \ \ \ \ \ \ \ \ \ \ \ \ \ \ \ \ \ \ \ \
\ \ \ \ \ \ \ \ \ \ \ \ \ \ \ \ \ \ \
\end{center}
\end{table}

\newpage
\pagestyle{full}
\begin{table}
\begin{center}
\caption{Reg 1 patches.}
\bigskip
\begin{tabular}{*{4}{c}}
\hline\hline
Patch id & ${\mit\Delta} \alpha $ & ${\mit\Delta} \delta $ & Area \\
& ('') & ('') & (arcmin$^{2}$) \\ \hline
1 &  397.5 &  -422.5  &    1.98 \\
2 &   249.9 &  -277.1  &    1.85 \\
3 &  134.8 &  -221.6  &    2.51 \\
4 &   140.6 &   -89.8  &    2.54 \\
5 &  -139.0 &   205.3  &    2.25 \\
6 &  -248.3 &   324.9  &    1.71 \\ \hline
\end{tabular}
\label {Table 2}
\end{center}
\end{table}

\newpage
\begin{table}
\begin{center}
\caption{Reg2 patches.}
\bigskip
\begin{tabular}{*{4}{c}}
\hline\hline
Patch id & ${\mit\Delta} \alpha$ & ${\mit\Delta} \delta$ & Area \\
& ('') & ('') & (arcmin$^{2}$) \\ \hline
1 &  129.1 &  -437.6  &    1.83 \\
2 &  -116.6 & -407.9  &    2.40 \\
3 &  -261.6 & -409.6 &    2.30 \\
4 &  -125.5 & -126.5 &    1.95 \\
5 &  -198.1 & -248.7 &    1.80 \\
6 &  -333.9 & -244.3 &    2.29 \\
7 &  -224.5 &  -29.9 &    1.84 \\
8 &  -381.1 & -122.8 &    2.37 \\
9 &  -357.5 &   95.6 &    2.71 \\
10 &  -374.9 &  273.2 &    4.93 \\
11 &  -435.2 &  123.9 &    2.69 \\
12 &   460.7 & -314.5 &    2.38 \\
13 &   351.5 & -183.6 &    2.49 \\
14 &   424.6 &   -2.9 &    3.26 \\
15 &   458.0 & -146.5 &    1.98 \\
16 &   220.0 &   57.3 &    1.10 \\
17 &   306.3 &  225.3 &    2.64 \\
18 &   470.7 &  128.0 &    2.05 \\
19 &   340.3 &  293.9 &    2.61 \\
20 &   184.2 &  302.4 &    2.73 \\
21 &   -43.6 &  378.2 &    2.69 \\
22 &   474.3 &  228.1 &    2.21  \\ \hline
\end{tabular}
\label {Table 3}
\end{center}
\end{table}

\newpage
\begin{table}
\begin{center}
\caption{Expected number of foreground $M$ giants towards BW
in the grism survey of FW87 for 2 different models of the star distribution,
assuming distances to the bulge of 4 and 6 kpc. GJ model (Garwood and Jones
1987) and Kent model (Kent et al. 1991). }
\bigskip
\begin{tabular}{*{6}{c}}
\hline\hline
MAGNITUDE & \multicolumn{2}{c}{GJ MODEL} & \multicolumn{2}{c}{KENT MODEL} &
NUMBER OBSERVED \\ \cline{2-5}
& $4$ kpc & $6$ kpc & $4$ kpc & $6$ kpc &  \\ \hline
4.25 & 0.10 & 0.10 & 0.07 & 0.07 & 0 \\
4.75 & 0.20 & 0.22 & 0.13 & 0.14 & 0 \\
5.25 & 0.34 & 0.47 & 0.22 & 0.27 & 0 \\
5.75 & 0.58 & 0.77 & 0.35 & 0.43 & 1 \\
6.25 & 0.95 & 1.47 & 0.55 & 0.76 & 1 \\
6.75 & 1.05 & 2.60 & 0.61 & 1.20 & 4 \\
7.25 & 1.58 & 4.20 & 0.88 & 1.80 & 14 \\
7.75 & 1.70 & 3.70 & 0.92 & 1.70 & 15 \\
8.25 & 1.44 & 6.30 & 0.77 & 2.60 & 10 \\
8.75 & 1.57 & 6.50 & 0.79 & 2.50 & 14 \\ \hline
\end{tabular}
\label {Table 4}
\end{center}
\end{table}

\newpage
\begin{table}
\begin{center}
\caption{Photometry of stars in PGC with $K_{0} \leq 5$.}
\bigskip
\begin{tabular}{*{7}{c}}
\hline\hline
ID & ${\mit\Delta} \alpha $ & ${\mit\Delta} \delta $ & $K$ & $(H-K)$ & $K_{0}$
& $(H-K)_{0}$  \\
& ('') & ('') & \\ \hline
1  &    -80.4 &   -489.3 &   9.58  &  4.53  &  3.92 &   0.75 \\
2  &   133.4  &   -243.9  &   9.41  &  3.92 &   4.56 &   0.68 \\
3  &   255.0  &    379.9  &  10.17  &  4.36 &   4.64 &   0.67 \\
4  &    98.3  &   -284.7  &  10.93  &  4.79 &   4.74 &   0.66 \\
5  &   286.3  &   -439.9  &   7.73  &  2.60 &   4.81 &   0.65 \\
6  &  -228.8  &   -106.4  &  10.83  &  4.62 &   4.87 &   0.65 \\
7  &   118.0  &    390.8  &   7.86  &  2.63 &   4.88 &   0.65 \\
8  &   406.3  &   -127.2  &   7.75  &  2.55 &   4.89 &   0.65 \\
9  &   134.9  &   -281.1 &  10.82  &  4.59 &   4.90 &   0.64 \\
10 &   -226.9 &      57.8  &   8.12 &   2.79 &   4.90 &   0.64 \\
11 &   -446.0 &     169.9  &   8.85 &   3.26 &   4.91 &   0.64 \\
12 &   -507.5 &     -16.3  &   8.77 &   3.21 &   4.93 &   0.64 \\
13 &   -284.5 &     113.6  &   8.61 &   3.09 &   4.94 &   0.64 \\
14 &    137.5 &     209.2  &   8.00 &   2.67 &   4.95 &   0.64 \\
15 &  -157.7  &    136.4  &   7.55  &  2.36 &   4.96 &   0.64 \\
16 &    438.1 &     290.8  &   8.98 &   3.30 &   4.98 &   0.64 \\ \hline

\end{tabular}

\label {Table 5}
\end{center}
\end{table}

\setcounter{figure}{2}
\begin{figure}
\begin{center}
\leavevmode
\hbox{%
\epsfxsize=7.25in
\epsffile{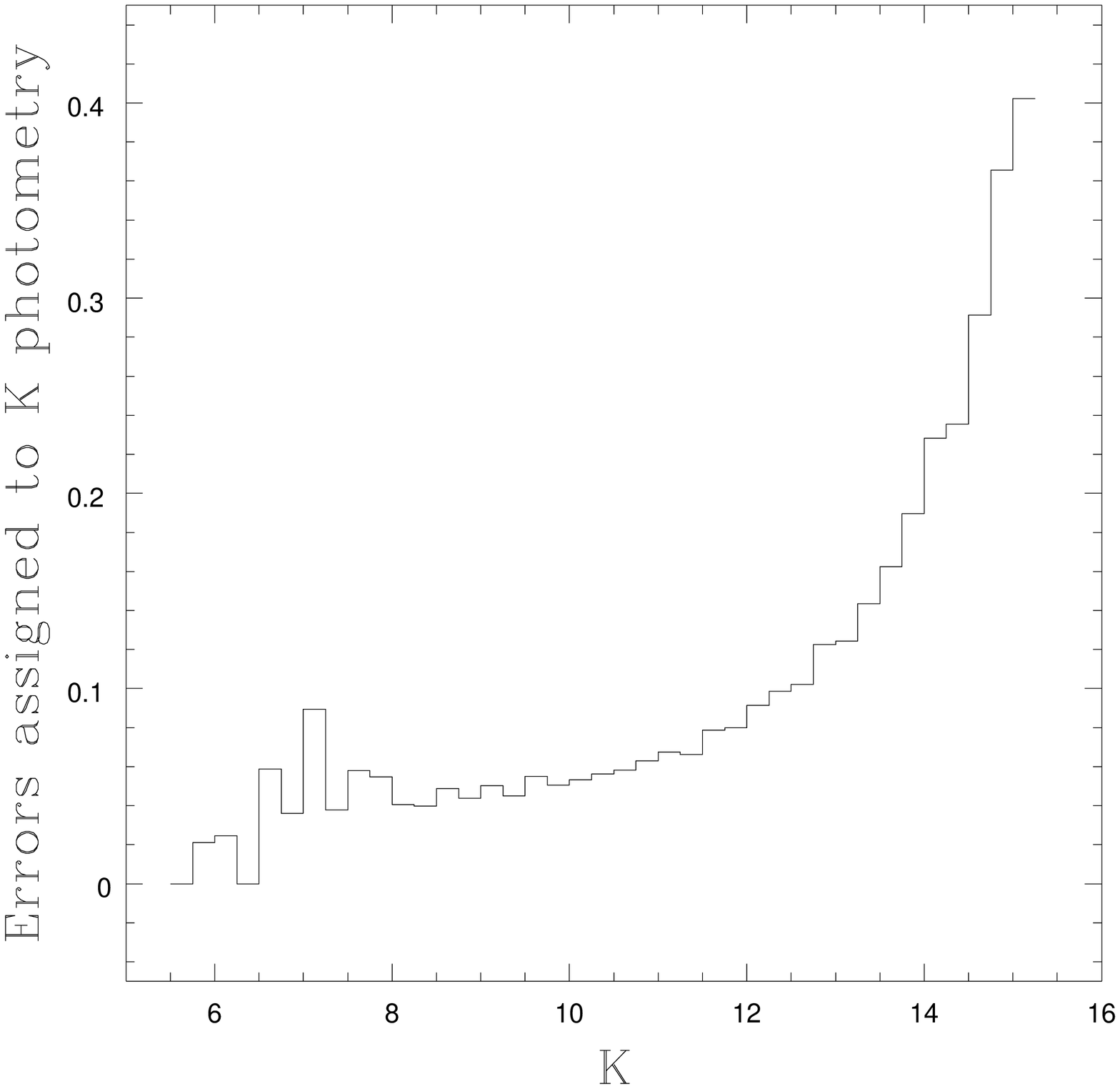}}
\caption{Errors in the K Photometry.}
\end{center}
\end{figure}

\begin{figure}
\begin{center}
\leavevmode
\hbox{%
\epsfxsize=7.25in
\epsffile{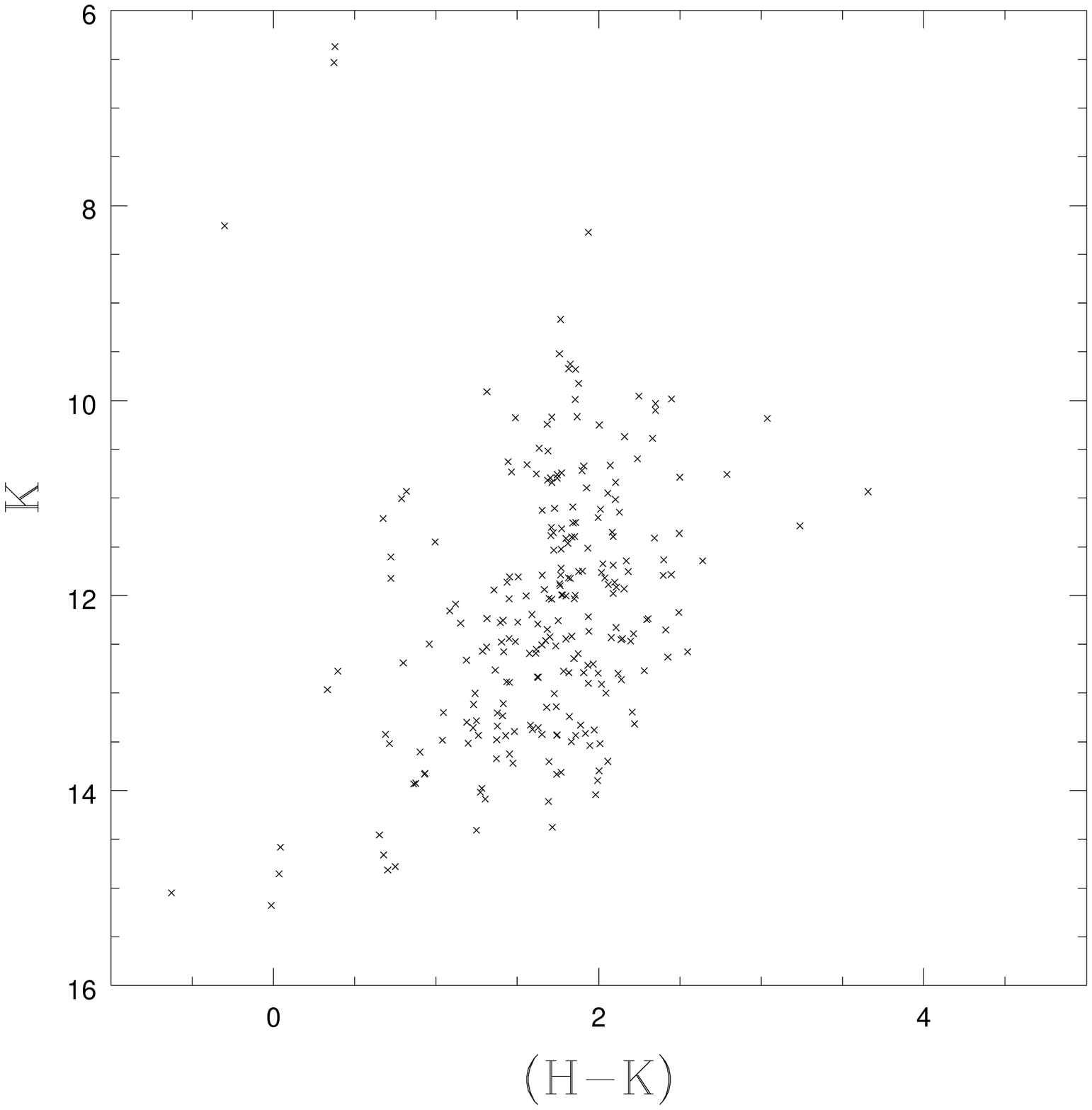}}
\caption{Color-mag diagram of patch 1 of Reg1.}

\end{center}
\end{figure}

\begin{figure}
\begin{center}
\leavevmode
\hbox{%
\epsfxsize=7.10in
\epsffile{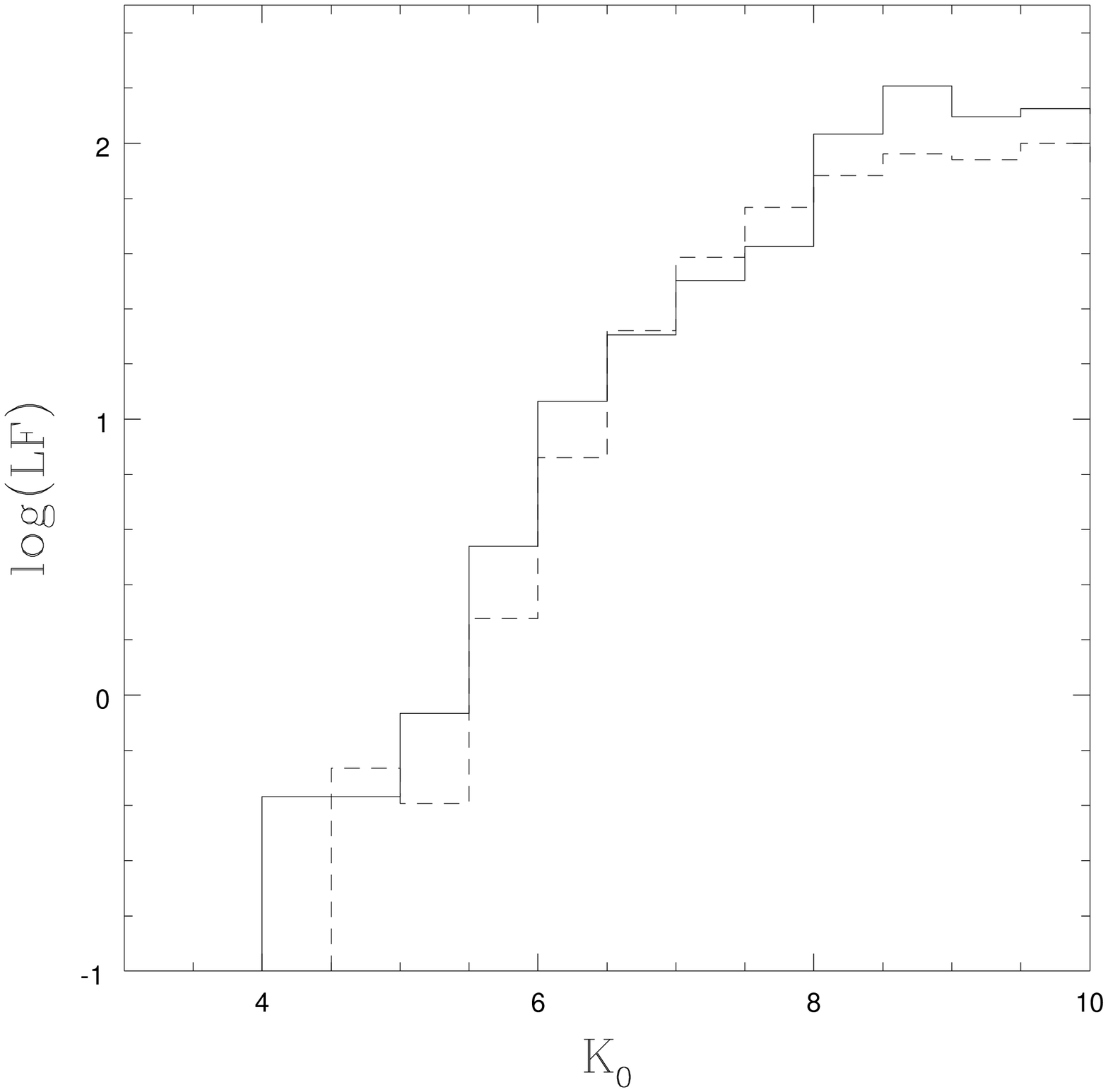}}
\caption{Comparison of the LFs from two different regions of PGC. Reg1 (solid
line) and Reg2 (dashed line). The two LFs are identical within
Poisson errors.}
\end{center}
\end{figure}

\begin{figure}
\begin{center}
\leavevmode
\hbox{%
\epsfxsize=6.75in
\epsffile{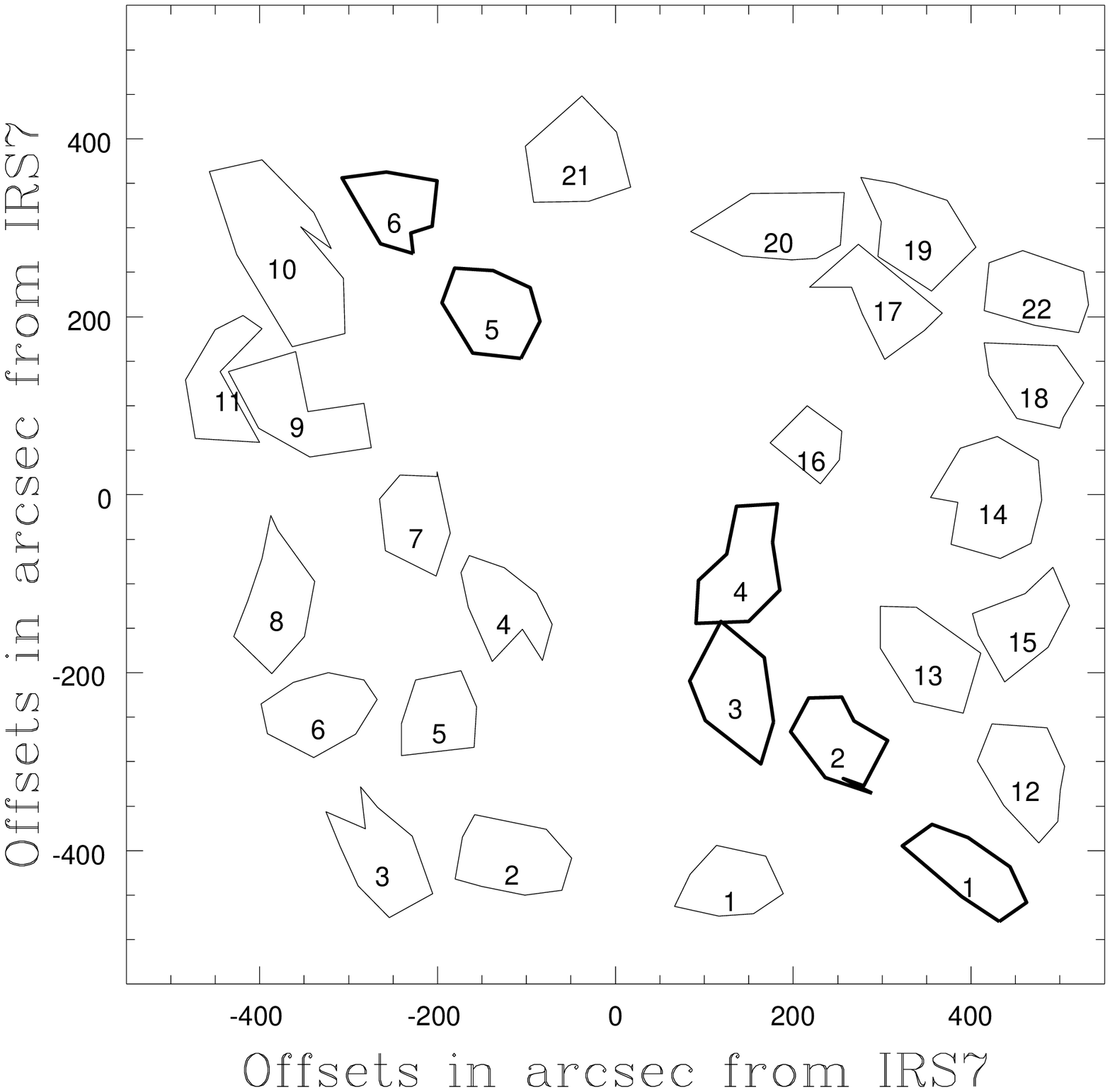}}
\caption{The 6 regions comprising Reg1(thick lines) and the 22 regions
comprising Reg2 (ordinary lines). North is up and east is left. Reg1 patches
lie in the disk like feature in the K mosaic, while the Reg2 regions
are outside it.}
\end{center}
\end{figure}

\begin{figure}
\begin{center}
\leavevmode
\hbox{%
\epsfxsize=7.25in
\epsffile{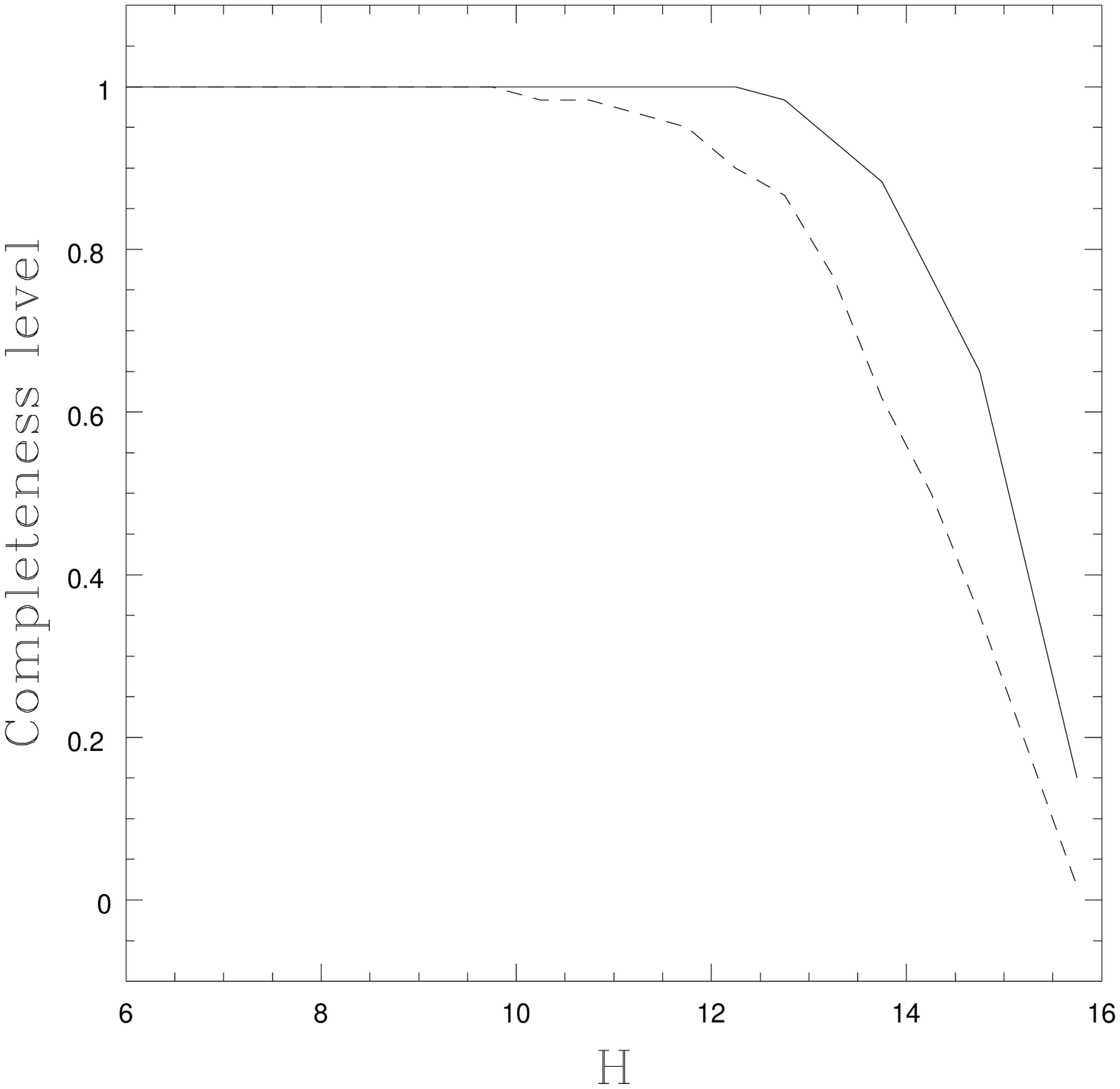}}
\caption{Completeness level as a function of $H$ for the total star list
 (solid line) and the DAOPHOT star list (dashed line).}
\end{center}
\end{figure}

\begin{figure}
\begin{center}
\leavevmode
\hbox{%
\epsfxsize=7.0in
\epsffile{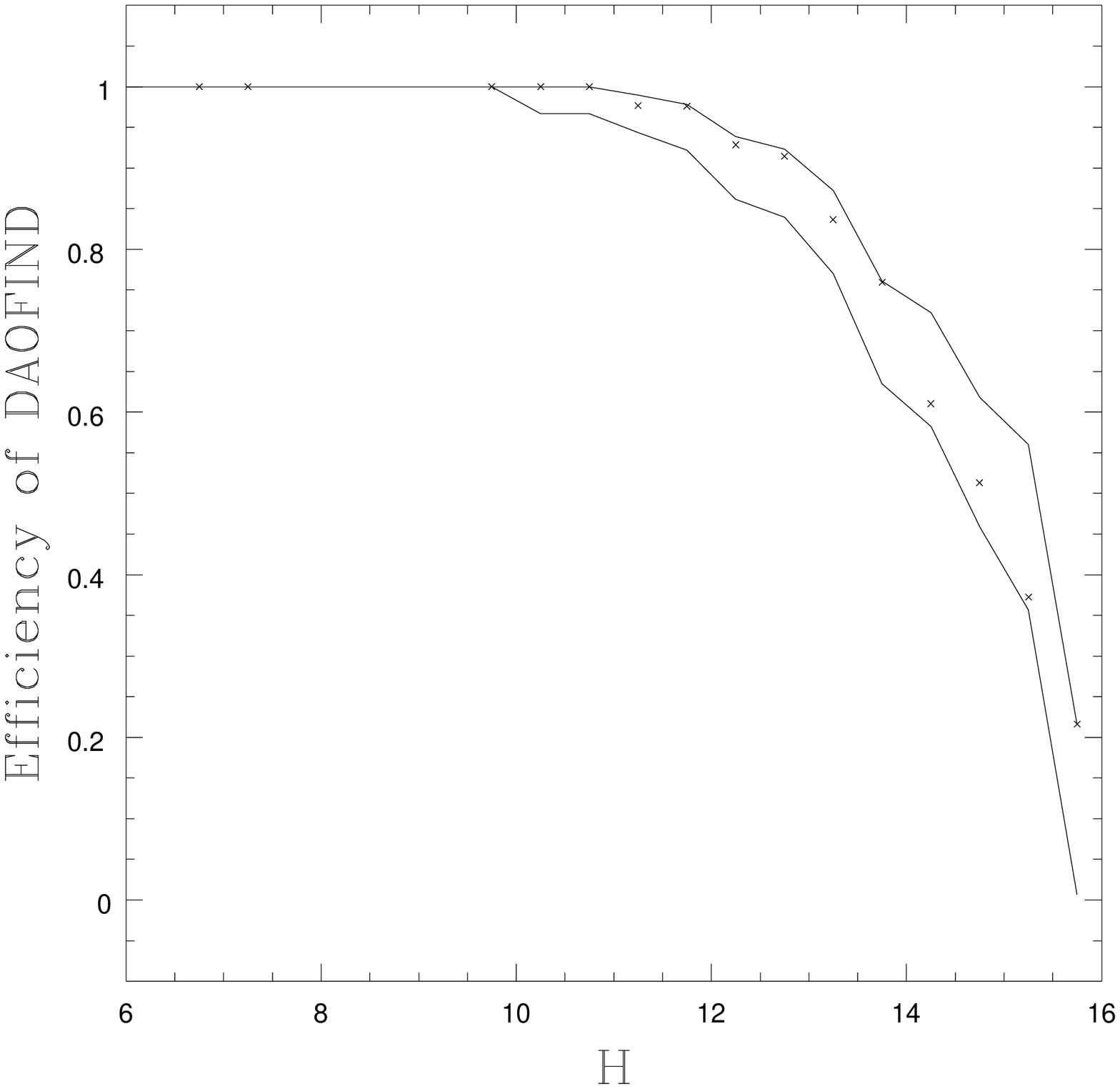}}
\caption{The two solid curves enclose the $1 \sigma$ bound of the DAOPHOT
detection fraction determined from the artificial star tests. The points are
the fraction for the sample star list.}
\end{center}
\end{figure}

\begin{figure}
\begin{center}
\leavevmode
\hbox{%
\epsfxsize=7.10 in
\epsffile{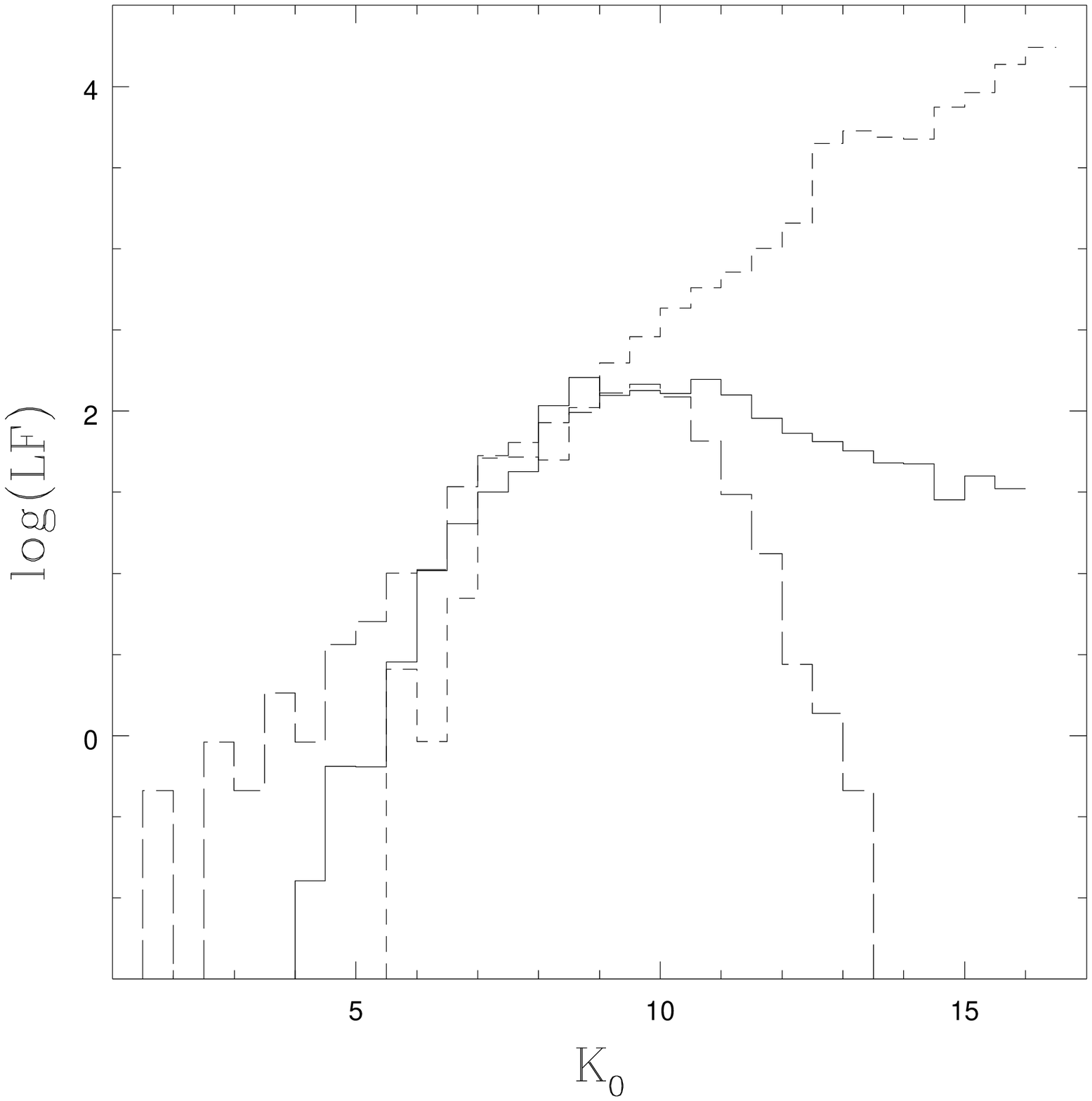}}
\caption{LFs of PGC (solid line), GC (long dashed line) and BW
(short dashed  line). For $K_{0} < 6.5$, the PGC LF is constructed from
both Reg1 and Reg2. }
\end{center}
\end{figure}

\end{document}